\documentclass[number]{elsarticle}
\usepackage{graphicx,longtable}
\usepackage{amsmath}
\usepackage{natbib}
\usepackage{aas_macros}
\pdfoutput=1
\usepackage{hyperref}

\begin{document}

\title{NSE abundance data}

\author{Andrzej Odrzywolek}
\ead{odrzywolek@th.if.uj.edu.pl}
\ead[url]{http://www.ribes.if.uj.edu.pl/}
\address{
M. Smoluchowski Institute of Physics\\ 
Jagiellonian University\\  
Reymonta 4\\ 
30-059 Krakow\\ 
Poland
}

\date{\today}

\begin{abstract}

Novel method of calculating Nuclear Statistical Equilibrium is presented.
Basic equations are carefully solved using arbitrary precision arithmetic. Special
interpolation procedure is then used to retrieve all abundances using tabulated
results for neutrons and protons, together with basic nuclear data.
Proton and neutron abundance tables,
basic nuclear data and partition functions for nuclides used in calculations
are provided. Simple interpolation algorithm using pre-calculated
p and n abundances tabulated as a functions of kT, $\rho$ and $Y_e$ is outlined. 
Unique properties 
of this method are:
(1) ability to pick-up out of NSE selected nuclei only 
(2) computational time scaling linearly with
number of re-calculated abundances 
(3) relatively small amount of stored data: only two large tables
(4) slightly faster than solving NSE equations using traditional Newton-Raphson methods
for small networks (few tens of species); superior for huge (800-3000) networks
(5) do not require initial guess; works well on random input
(6) can tailored to specific application
(7) ability to use third-party NSE solvers to obtain fully compatible tables
(8) encapsulation of the NSE code for bug-free calculations.

Range of applications for this approach is possible: coverage test of traditional
NSE Newton-Raphson codes, generating starting values, code-to-code verification
and possible replacement of the old legacy procedures in supernova simulations.

\end{abstract}

\maketitle

\tableofcontents
\listoffigures 
\listoftables

\section{Brief introduction}

Main goal of the article is to provide new method of computing the 
Nuclear Statistical Equilibrium (NSE) abundances of the nuclear species. 
Vast range of conditions can be analyzed: $Y_e=0.0\ldots1.0$,
$\rho=10^2 \ldots 10^{13}$ g/cc and $T=2 \times 10^9 \ldots 10^{11}$ K covering almost any
astrophysical situation imaginable. 
While this article do not concentrate on particular target object, these results are useful
for study of pre-supernova stars after Si burning \cite{OdrzywolekHegerActa}, thermonuclear
supernovae \cite{2010arXiv1006.0490O}, core-collapse \cite{2010arXiv1007.0463G}, 
and protoneutron stars \cite{2010arXiv1002.3854A}.

We calculate NSE abundances using reliable arbitrary precision arithmetic
approach. Tables of pre-calculated proton ($X_p$)
and neutron ($X_n$) abundances as functions of the thermodynamic conditions
defined by the triad: $\rho, T, Y_e$ are stored. Recovering of the remaining 
several hundred abundances from these two tables is non-trivial task.
Detailed description of the working procedure used to calculate
NSE abundances is provided. Algorithm is fast thanks to use of pre-tabulated
$X_{p,n}$. It has unique ability to pick up out of NSE ensemble
only species of interest and other features. Computational time scales {\em linearly} with 
number of required nuclides.

\section{NSE \label{NSE}}

\subsection{Basic equations}

      Well-known equations for the ensemble of $N_{iso}+1$ nuclei in thermal 
      equilibrium \cite{1994ApJS...91..389A, 2009ADNDT..95...96S} are: 
      \begin{subequations}
      \label{nse-equations}
      \begin{equation}
      \label{nse-equations-1}  
           \sum_{k=0}^{N_{iso}} X_k = 1
      \end{equation}
      \begin{equation}
      \label{nse-equations-2}
            \sum_{k=0}^{N_{iso}} \frac{Z_k}{A_k} \, X_k = Y_e
      \end{equation}
      \end{subequations}
      where abundance $X_k$ for $k$-th nuclei with atomic number $Z_k$
      and mass number $A_k$ is:
      \begin{equation}
      \label{nse-abundances}
           X_k = \frac{1}{2}\, G_k(T) \,
           \left( \frac{1}{2} \rho N_A \lambda^3 \right)^{A_k-1}\,
           {A_k}^{5/2}\,X_n^{{A_k}-{Z_k}} X_p^{Z_k}\, e^{\frac{Q_k}{kT}}.
      \end{equation}
      Temperature dependent partition function for $k$-th nuclei is given by:
      \begin{equation}
      \label{G_kT}
      G_k(T) = \sum_{i=0}^{i_{max}} (2 J_{ik} +1)  e^{-\frac{E_{ik}}{kT}}
      \end{equation} 
      where summation is over all known excited states (numbered by the index i) 
      of the k-th nucleus; $J_{ik}$ and
      $E_{ik}$ are the spin and the excitation energy, respectively. $Q_k$
      is the binding energy, $\rho, T$ - density and temperature of the plasma,
      $N_A$ is the Avogadro number and $k$ - Boltzmann constant. Thermal de'Broglie
      wavelength used in eq.~\eqref{nse-abundances} is:
      \begin{equation}
      \label{thermal-debroglie}
      \lambda = \frac{h}{\sqrt{2 \pi m_{H} kT }} 
      \end{equation}
      where $m_{H}$ is the mass of the hydrogen atom and $h$ denotes Planck's constant.

\begin{figure}
\includegraphics[width=\textwidth]{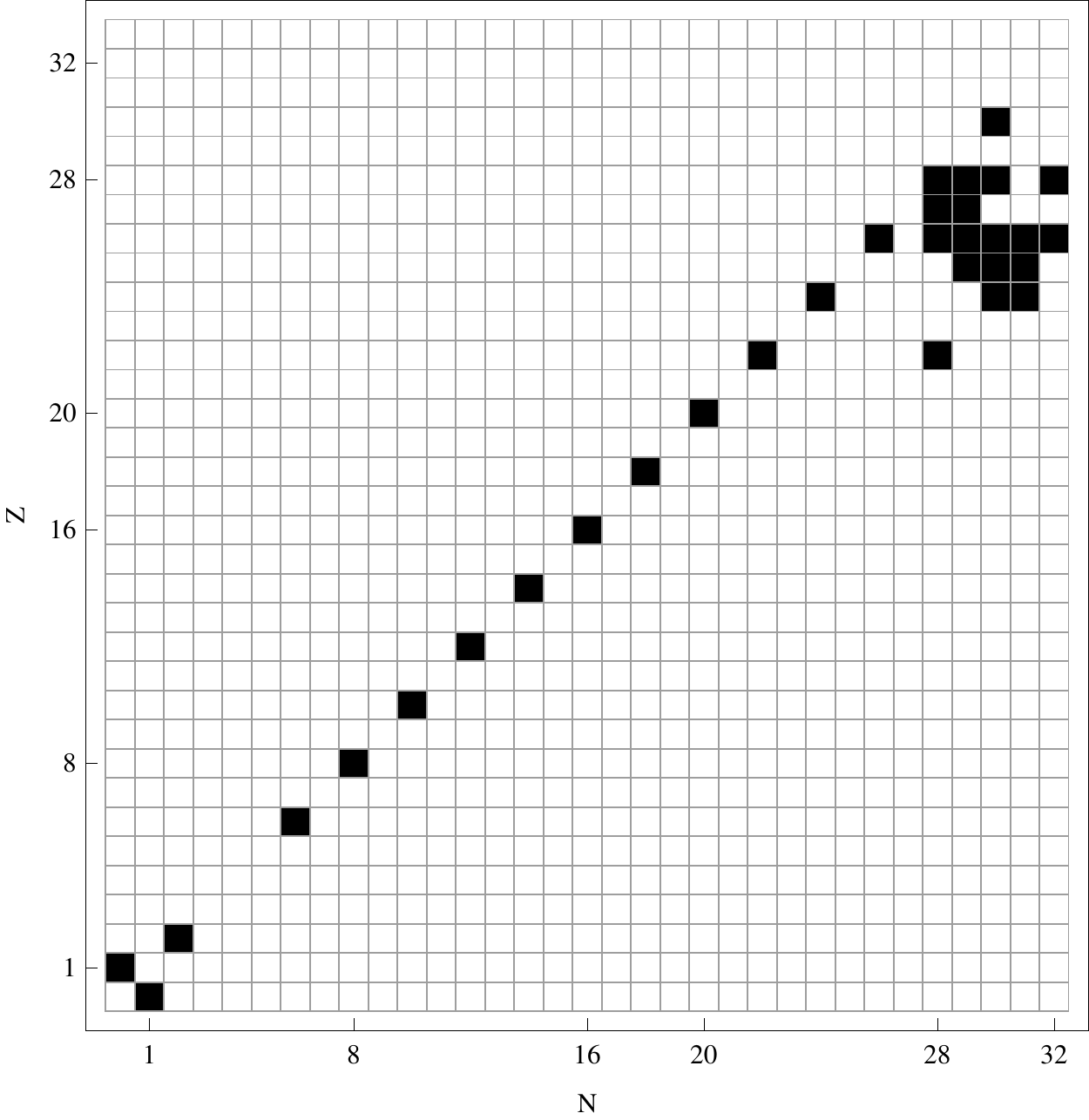}
\caption[Nuclides included in NSE calculations.]
{
\label{chart} 
Nuclides included in NSE calculations.
Much larger (800 nuclear species) networks were also tested.
}
\end{figure}

\begin{figure}
\includegraphics[width=\textwidth]{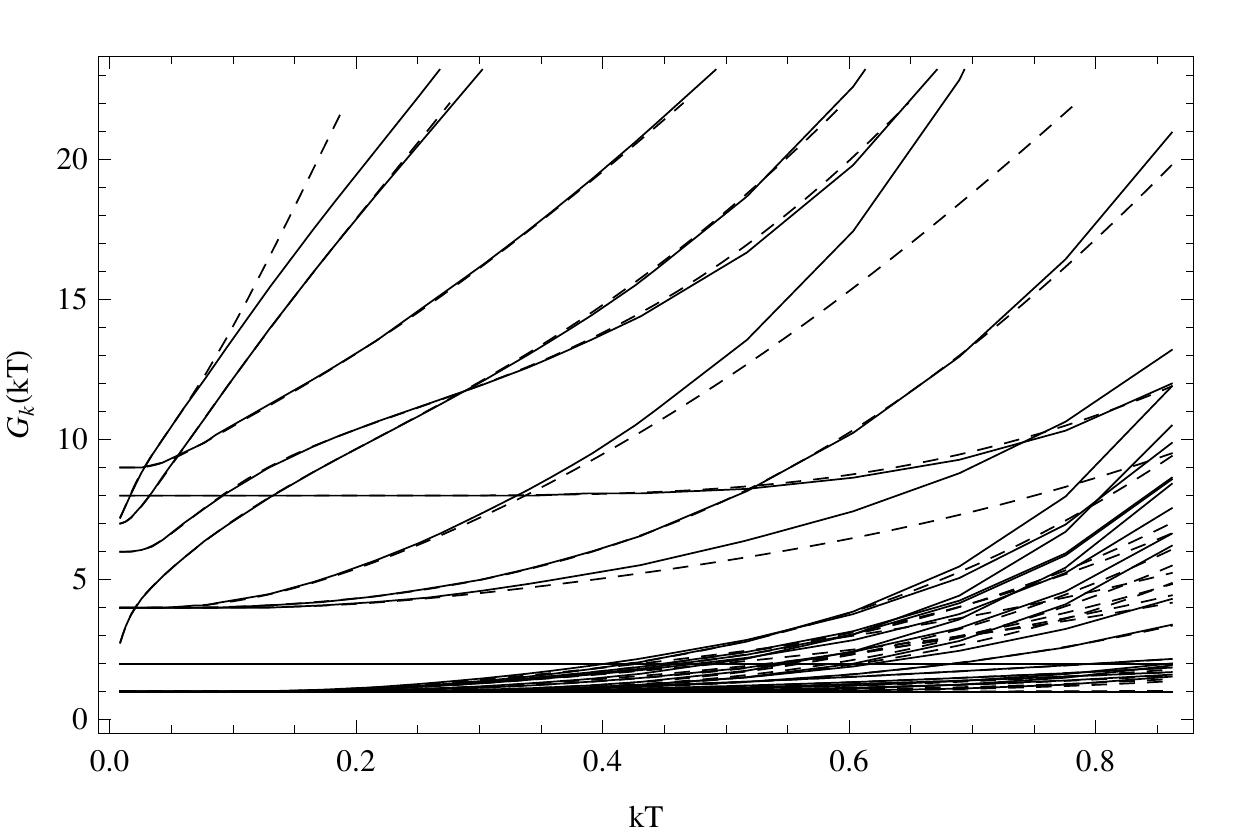}
\caption[Comparison of the partition functions.]{\label{partfun} Comparison of the partition functions derived directly
using \eqref{G_kT} and database \cite{MATHEMATICA_IsotopeData} (solid lines),
and those from Hix \& Thielemann code \cite{1996ApJ...460..869H,1999ApJ...511..862H}
(dashed lines).
}
\end{figure}

Partition function has been calculated directly from nuclear database
using \eqref{G_kT}. Missing
spins were assumed to be equal zero. For uncertain lowest possible was used.
Results are in good agreement with data used 
in \cite{1996ApJ...460..869H,1999ApJ...511..862H}, cf. Fig.~\ref{partfun}.

\subsection{Limitations of the Newton-Raphson NSE solvers}

  NSE equations, from mathematical point of view, form the system of two 
large high order polynomial equations (polynomial system) for unknown proton $X_p$
and neutron $X_n$ abundances. System is solved
numerically using two-dimensional Newton-Raphson 
technique\footnote{In principle polynomial system might be reduced using Groebner
basis methods, especially over rational field. In practice ensemble including
protons, neutrons, $^4$He and single heavy nuclei can be solved, but additional components
cause Groebner basis algorithms to fail in sense of computational time: no result is returned
in a period of several hours.}.
Due to large integer powers and other factors this approach is prone to numerous convergence
problems. While in ''normal'' situation (typical thermodynamic parameters,
good initial guess, standard selection of species) convergence 
of codes using machine floating point arithmetic
is amazingly fast, failures are inevitable.       
Limited numerical precision might be problematic issue. 
This forces programmers to include multi-level fail-safe procedures. They are by many orders
of magnitude slower, and not guaranteed to converge. 
Careful programming with proper handling of round-off errors 
is required to get correct results, leading to additional complications.
Due to problems with numerical precision and unpredictable iteration numbers, 
rapidly growing with number of species and for low temperatures, procedures are long, 
complicated and hard to parallelize.

Moreover, even if we are interested in abundance of single nucleus entire system \eqref{nse-equations}
has to be solved. Such a situation is typical for neutrino spectrum calculations, as usually much more nuclear 
species are included in NSE than those with known neutrino emission rates. Usually very few 
of them contribute at non-negligible level, e.g. $p$, $^{56}$Ni and $^{55}$Co for $\nu_e$ emission at $Y_e=0.5$. 
Large part of $kT-\rho-Y_e$ space is completely
dominated by processes involving neutrons and protons only.   
In the course of the research we have faced this problem. In recent article 
\cite{2010arXiv1006.0490O} NSE ensemble included 800 nuclides 
while FFN tables used include only 189 of them. Interpolation of the pre-calculated results 
has been found to be optimal solution. Similar problem arise in core-collapse supernova 
simulations. Depending on temperature, NSE or full reaction network is solved. Again, 
NSE can be computed for larger ensemble, but due to limitations of network ODE solvers
only fraction of species is traced.

\subsection{Interpolation algorithm \label{algorithm} }

To handle results of the NSE calculations efficiently, interpolation seems to be wrong 
solution. Naively,
one might try to interpolate stored proton $X_p$ and neutron $X_n$ abundances obtained 
from Eqns.~\eqref{nse-equations},
and get $X_k$ from \eqref{nse-abundances}. Unfortunately, this does not work. 
Even a very small inaccuracy
in $X_n$ or $X_p$ produces enormous errors\footnote{
This relative error can be estimated as: $\delta (X_n^N X_p^Z) \sim A\, 2^{A} \delta X$, where $\delta X$ is typical
relative error of $X_n$ ($X_p$) and $A$ is mass number. For $A\sim60$ amplification of relative error
might be as large as $10^{18}$ (!) for $X_n \sim X_p \sim 0.5$.
} 
in $X_k$ due to large ($\sim A$) integer powers in \eqref{nse-abundances}. 
Another ,,brute force'' method is tabulation of every $X_k$. 
This might be useful if a few  out of NSE species are of interest. This is also the fastest
approach. However, for larger number of species amount of stored 
data becomes very large: tens or hundreds of tables like Table~\ref{pn_table} instead of two. Fortunately,  
we found a compromise, which successfully combines both ideas.
Inability to get accurate abundances using interpolated $X_n, X_p$ does not include grid 
points, as they can be stored with accuracy 
equal to the machine precision, or even better if required.
First,  we calculate abundance of selected species $X_k$ at grid points neighboring 
given ($\rho, T, Y_e$)
point.  Next, we interpolate using computed $X_k$'s. Only proton $X_p$ and neutron $X_p$ abundances need to be tabulated,
but more (using formula \eqref{nse-abundances} at 8 corners of a cube)
computational time is 
required compared to interpolation of stored $X_k$ values for all nuclei.
Additionally, partition function $G_k(T)$, atomic and mass numbers $Z_k, A_k$ and binding energy $Q_k$ 
has to be stored for all nuclei to use \eqref{nse-abundances}.  Using (tri)linear interpolation
eqns. \eqref{nse-equations} are fulfilled automatically up to original solving accuracy.

We still have to solve \eqref{nse-equations} to generate $X_p$ and $X_p$ tables. Any method
e.g. existing codes \cite{FXT_NSE}, pre-calculated results or a web service \cite{webnucleo} 
may be used in this purpose.
Because efficiency and speed of the code is not of primary importance if one use  
interpolating scheme,
Eqns.~(\ref{nse-equations-1}, \ref{nse-equations-2}) has been solved  numerically using 
MATHEMATICA code.\footnote{Entire code \cite{PSNS}
has approx. 100 lines including database loading, writing C headers, and solving \eqref{nse-equations} with 
arbitrary precision. Code is slow compared to FORTRAN equivalents, 
a price paid for arbitrary precision. This is not an important issue, 
as all we want is to generate tables. We do it once, in parallel if required. 
Later we use interpolators, which are
very fast, even compared to codes using hardware floats.} 
Integrated MATHEMATICA \cite{MATHEMATICA_IsotopeData} database has been used,
including excited states
and spins. This let us to calculate temperature dependent partition function. 
Measured excited states
were used if present in database, otherwise neglected. Third party partition functions can
be used as well.      
No Coulomb and screening corrections were applied.

Proton and neutron abundances are then tabulated as a functions 
of temperature, density and electron fraction.
NSE results are checked against available codes/results 
\cite{FXT_NSE, webnucleo, 2008ApJ...685L.129S} with good agreement.
      
\subsection{Brief discussion of NSE results}

\begin{table}
\caption[Minimum number of nuclides.]{\label{Xmin_tbl} Minimum number of nuclides required to compute {\em all} abundances above $X_{min}$.}
  \begin{tabular}{c|ccc c}
  $X_{min}$  &   Z &  A & niso &   Last included nuclide   \\
  $10^{-1}$  &  28 & 56 &  562 & $^{56}$Ni   \\
  $10^{-2}$  &  28 & 57 &  563 & $^{57}$Ni   \\
  $10^{-4}$  &  29 & 59 &  592 & $^{59}$Cu   \\
  $10^{-5}$  &  30 & 60 &  620 & $^{60}$Zn   \\
  $10^{-6}$  &  30 & 61 &  621 & $^{61}$Zn   \\
  $10^{-7}$  &  30 & 63 &  623 & $^{63}$Zn   \\
  $10^{-8}$  &  31 & 63 &  651 & $^{63}$Ga   \\
  $10^{-9}$  &  31 & 65 &  653 & $^{65}$Ga   \\
  $10^{-10}$ &  32 & 66 &  683 & $^{66}$Ge   \\
  $10^{-12}$ &  32 & 68 &  685 & $^{68}$Ge   \\         
  $10^{-20}$ &  36 & 75 &  807 & $^{75}$Kr   \\
  $10^{-30}$ &  41 & 87 &  970 & $^{87}$Nb
  \end{tabular}
\end{table}

Determination of NSE abundances is crucial for many applications, including nucleosynthesis, neutrino
emission, nuclear energy generation and equation of state. Therefore we have made some tests to 
verify results and accuracy estimates.
Despite known physical issues (temperature-dependent partition function, 
Coulomb corrections \cite{2008ApJ...685L.129S}, screening \cite{1990ApJ...362..620I}) 
one of the most important factors is number and selection of species included in equations \eqref{nse-equations}.
Even a single one important nuclei missing in NSE ensemble may lead to radically different 
results. While inclusion of some species seems obvious (p, n , $^4$He, $^{56}$Ni, iron group)
further selection is more or less arbitrary.

\begin{figure}
  \includegraphics[width=\textwidth]{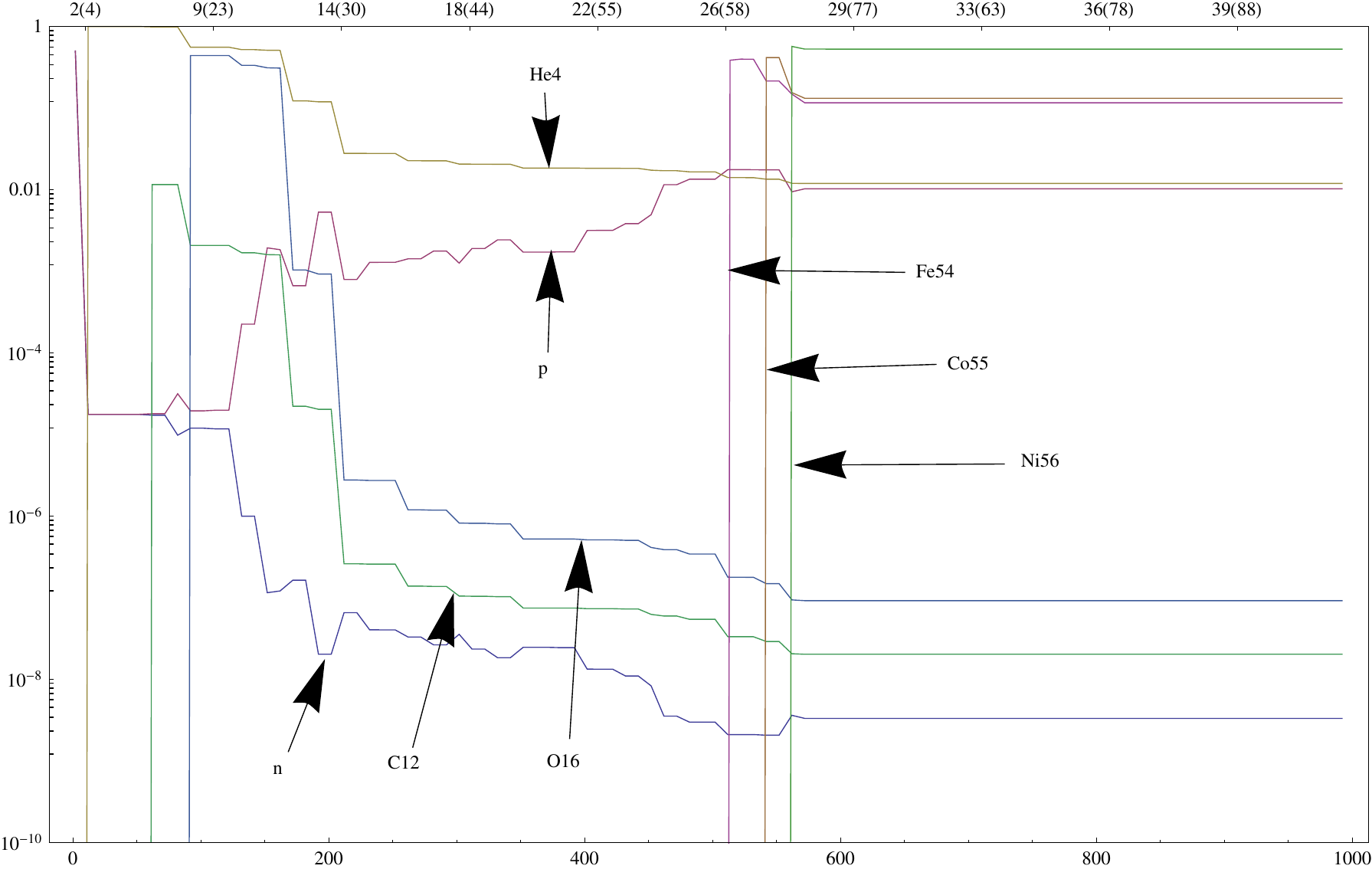}
  \caption{\label{NSE-numiso} NSE abundances as a function of the number of the nuclei involved in calculations
           for $kT=0.4$~MeV, $\rho=10^7$ g/cm$^3$ and $Y_e=0.5$.}
\end{figure}

To quantify problem I tried to answer the following question: what is the maximum required
atomic (Z) and mass number (A) to get solution including all species with abundance 
larger than prescribed $X_{min}$. Results are presented in Table~\ref{Xmin_tbl} and Fig.~\ref{NSE-numiso}. For example,
from Table~\ref{Xmin_tbl}, if we do not want to miss any of species with abundance above 
e.g. $10^{-6}$, we need at least nuclides up to $^{61}$Zn. Nuclei in Fig.~\ref{NSE-numiso} are ordered
according to \cite{MATHEMATICA_IsotopeData}; approximate Z and A are included as a tick marks for a top axis.
This estimate gives an upper limit for number of required nuclei. To get true minimal number of nuclides
required to get all species above assumed accuracy one have to consider all subsets for entire considered
$kT-\rho-Y_e$ space. Number of subsets, given by the Bell number $B_{N_{iso}}$ is 
very large.
Therefore, rigorous selection of species is impossible for large sets, and the safest thing to do is to
use estimates given by Table~\ref{Xmin_tbl} or consider all nuclei available \cite{webnucleo}.
In practice however, other factors decide, e.g. limited computational resources in supernova simulations.

From Fig.~\ref{NSE-numiso} we can conclude that the most primitive NSE including p and n only is not useful, 
maybe except for very high temperatures, cf.~Fig.~\ref{NSE_vs_kT}. Inclusion
of the alpha particle extends applications to lower temperatures but usually p and n abundances are
wrong by few orders of magnitude. To get correct abundances of p and n for lower temperatures
entire iron peak has to be included. $X_p$ and $X_n$  are rock-stable if all nuclei below 
Z=28, A=56 are included. This number might be seriously reduced if we focus on narrow
$Y_e$ range and exclude low mass (A=3..16) elements. Anyway, results in Table~\ref{Xmin_tbl} indicate, 
that no more than 1000 nuclei are required to get all abundances
above $10^{-30}$. While it is possible to solve NSE equation for more than 3000 nuclides \cite{webnucleo}, 
it does not change results significantly.

\subsection{NSE viewgraphs \label{viewgraphs} }

\begin{figure}
\includegraphics[width=\textwidth]{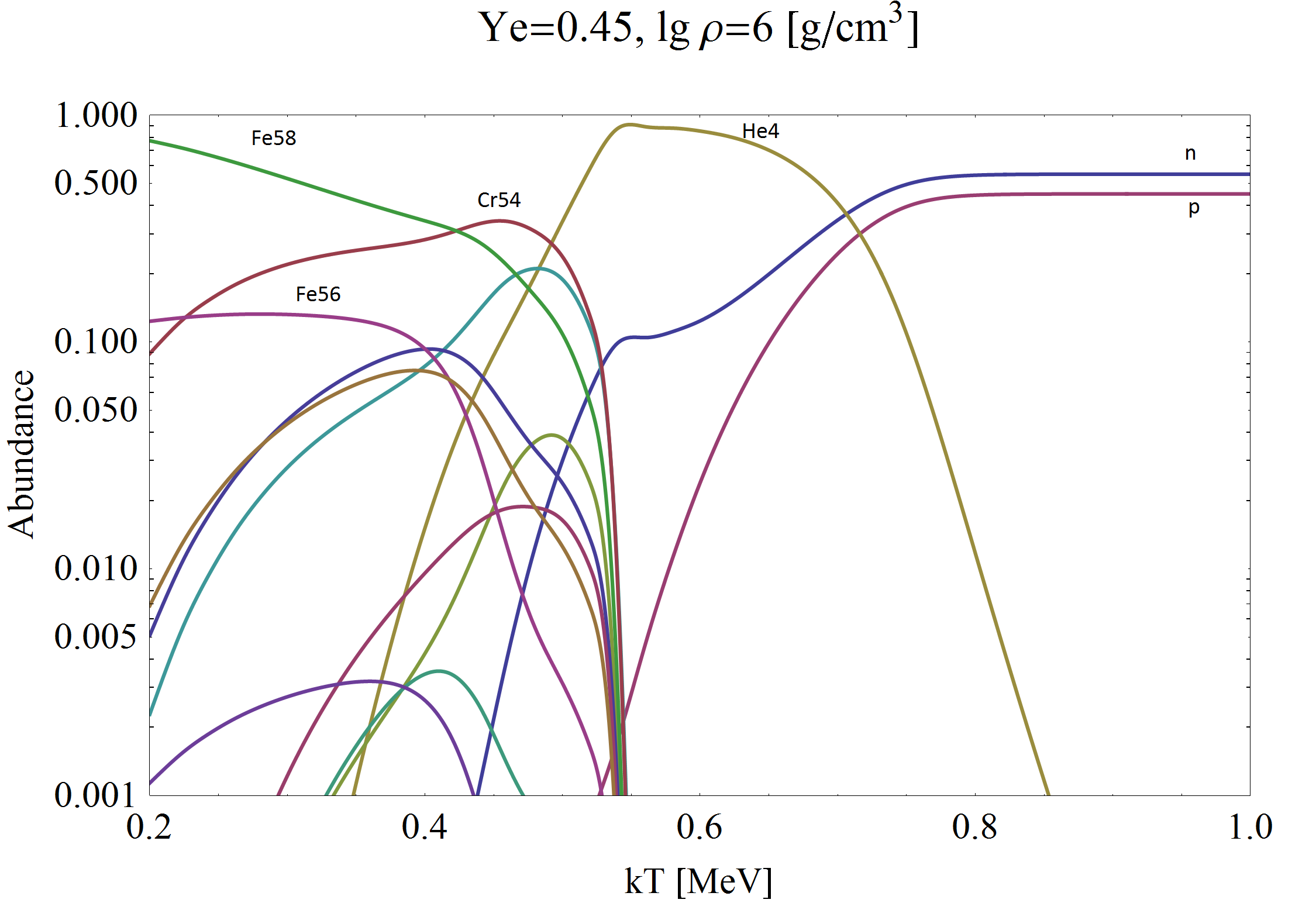}
\caption{\label{NSE_vs_kT} NSE abundance {\em versus} temperature.           }
\end{figure}

\begin{figure}
\includegraphics[width=\textwidth]{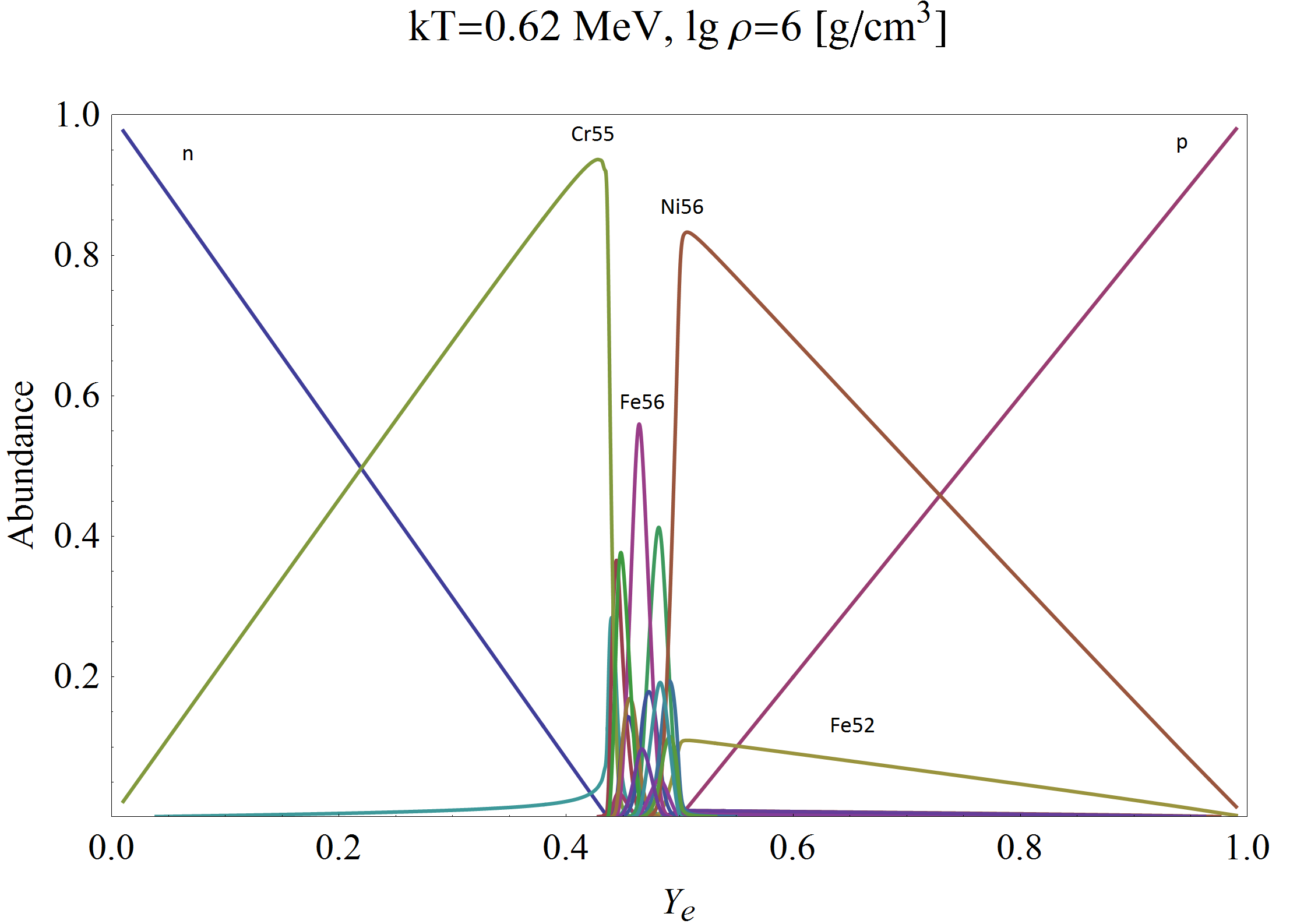}
\caption{\label{NSE_vs_Ye} NSE abundance {\em versus} electron fraction $Y_e$. }
\end{figure}

We discuss some properties of the NSE state for completeness.
For a very high 
temperatures\footnote{\label{footnote:dens} Actually, if we forget temperature dependent partition function, 
according to eq.~\eqref{nse-abundances}
solution depends on factor proportional to $\rho^2/kT^3$.
}
above $kT\simeq0.5$ MeV in Fig.~\ref{NSE_vs_kT} ($T_9\simeq$5.8) 
no bound nuclei exist
and we have a mixture of free neutrons and protons (Fig.~\ref{NSE_vs_kT}).
In this case solution of system \eqref{nse-equations} is:
$$
X_n = 1-Y_e, \qquad X_p = Y_e.
$$
If temperature decrease helium is being ''synthesized'' like in Big Bang 
nucleosynthesis. If temperature drops
further more below $kT\simeq0.35$~MeV ($T\simeq4 \times 10^9$~K) and 
thermodynamic conditions are maintained for long enough, heavy most bound nuclei
are preferred. Finally, cold catalyzed matter state is a pure (for $Y_e=0.45$) $^{58}$Fe, 
cf. Fig.~\ref{NSE_vs_kT}; for $Y_e=0.5$ it is $^{56}$Ni.
This is appealing physical picture.  Note
extremely strong $Y_e$ dependence of the NSE state (Fig.~\ref{NSE_vs_Ye}) for
 $0.35\!<\!Y_e\!<\!0.55$. $Y_e$ dependence
for large temperatures is trivial: smooth balance between p, n and $\alpha$ abundances.
The most interesting
is the temperature range where heavy nuclei dominate. Note that, for higher densities, 
temperature threshold 
for heavy nuclei formation moves to higher temperatures, see eq.~\eqref{nse-abundances}
and footnote \ref{footnote:dens}.

\begin{figure}
\includegraphics[width=\textwidth]{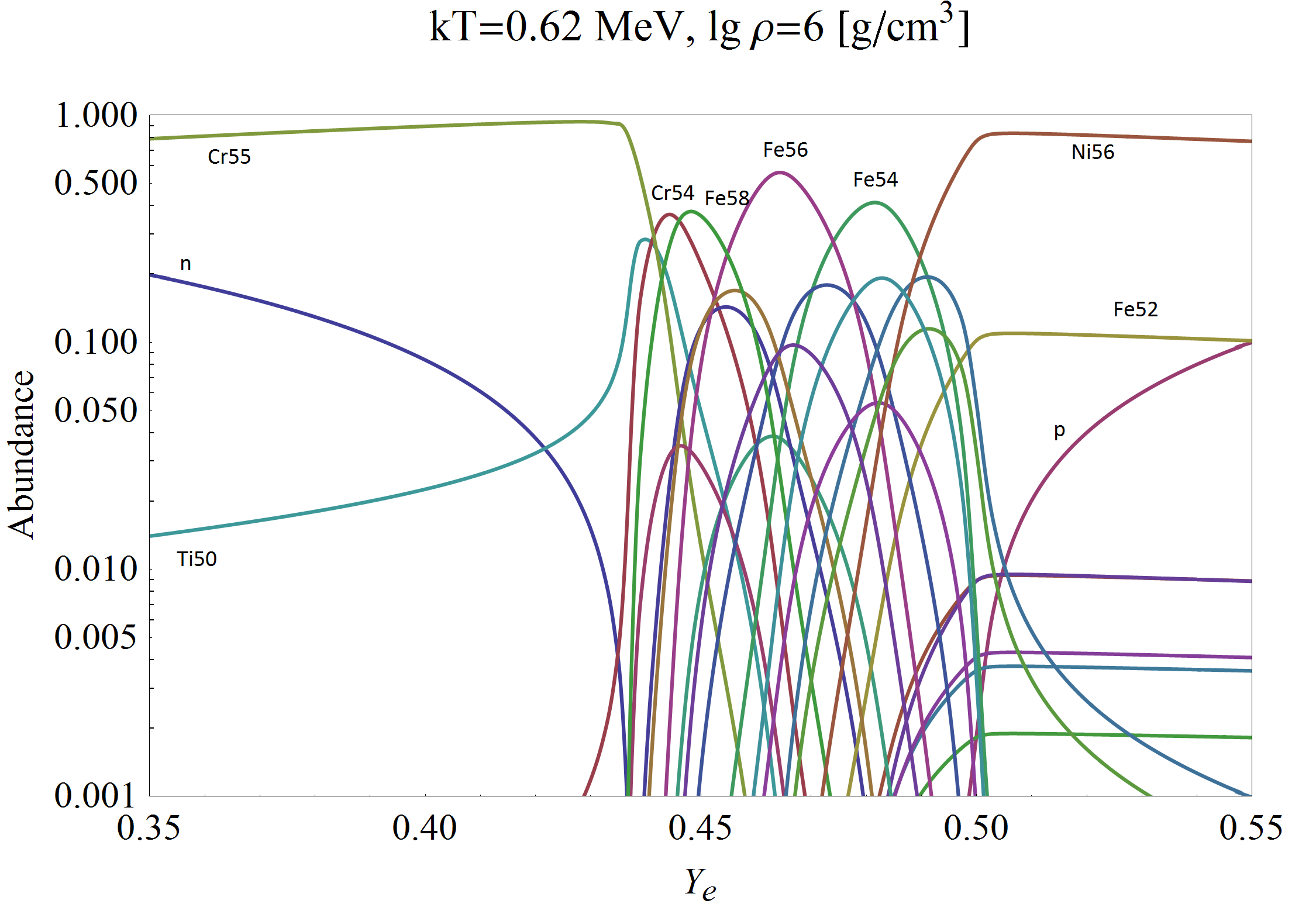}
\caption[Zoom of the Fig.~\ref{NSE_vs_Ye}.]
{\label{NSE_vs_Ye_zoom} 
Zoom of the Fig.~\ref{NSE_vs_Ye} into the most interesting range
of $Y_e=0.35\ldots0.55$.                       
}
\end{figure}

Striking feature of Fig.~\ref{NSE_vs_Ye} is a rapid variation 
of the abundances within range of $Y_e=0.35\ldots0.55$, cf. Fig.~\ref{NSE_vs_Ye_zoom}.  
NSE prefer nucleus with individual $Y_e^{(k)} \equiv Z_k/A_k$ as close as possible to 
$Y_e$ for entire thermodynamic ensemble. For example double magic nuclei 
$^{78}$Ni with largest known neutron 
excess\footnote{Neutron excess is equivalent to $Y_e$: $\eta=1-2 Y_e$.} 
(lowest $Y_e=28/78 \simeq0.36$) dominates for $Y_e<0.365$ 
(not included in example network)
until neutrons (with $Y_e=0$) take a lead.
For opposite side, $Y_e \gg 0.5$, 
protons are dominant\footnote{
Normally, for $Y_e\gg0.5$ protons dominate. But if $^3$Li
would exist, it should take the role of hydrogen under NSE conditions if 
density is high enough. This species is still present
in nuclide databases with atomic mass $3.030775$ and binding energy $2.2676$~MeV,
despite experimental detection \cite{1969PhRvL..23.1181W} has never been 
confirmed \cite{1987NuPhA.474....1T};
see also comments in ESNDF data at http://ie.lbl.gov/ensdf/.
}.

Note, that for $Y_e=0$ exact solution of NSE equations is $X_n=1$; for $Y_e=1$ we get $X_p=1$.
Abundances for intermediate values of $Y_e$ continuously approach these values 
for $Y_e \to 0$, and $Y_e \to 1$, cf. Fig.~\ref{NSE_vs_Ye}. 
Other abundances rapidly approach zero.

Rapid abundance variation has strong imprint on neutrino emission. 
For example, known for large electron capture rate $^{55}$Co has non-negligible abundance only in
narrow range of $Y_e=0.47\ldots0.5$.

\section{Proton and neutron NSE abundance tables: 
explanation and examples of use}

Dataset described and presented in the article is meant 
to be simple example of the methods used. It is tailored
to test against 32 isotope NSE solver used by Garching group,
based on serial code of Hix \& Thielemann \cite{1996ApJ...460..869H,1999ApJ...511..862H}.
In real application user should use larger tables for bigger networks
available online or generate (request from author) user-defined dataset
tailored to specific application.

Here we provide tables of the proton, $X_p$ and 
neutron, $X_n$ abundances, together with nuclear data required 
to calculate all remaining abundances, $X_k$. Additionally, list of nuclides is required,
including:
\begin{enumerate}
\item atomic and mass numbers
\item masses and binding energies
\item spins and excited states or, equivalently, tabulated
temperature-dependent partition function
\end{enumerate}

In Table~\ref{pn_table} we have provided:
in the first column temperature $kT$ in MeV,
second column include base 10 logarithm of the density in g/cm$^3$,
and third column electron fraction $Y_e$, i.e. number of electrons
divided by number of baryons. Fourth column include proton abundance
under NSE defined by the $kT, \rho$ and $Y_e$, and fifth column
contain neutron abundance.

\begin{table}
\caption{Symbols used in nuclide table, main Table~\ref{nuclei_table} 
begins on page \pageref{nuclei_table}.}
\begin{tabular}{c|c|c}
Col & Symbol & Description \\
\hline
\hline
1 & No     & position \\
2 & Symbol & Standard element symbol\\
3 & A      & Mass number   \\
4 & N      & Neutron number  \\
5 & Z      & Atomic number \\
6 & Q      & Binding energy per nucleon [MeV] \\
7 & $J_0$ & ground state spin (0 if not known)
\end{tabular}
\end{table}

\begin{table}
\caption[Symbols used in temperature dependent partition function table.]{
Symbols used in temperature dependent partition function table, 
main Table~\ref{partfun_table} 
begins on page \pageref{partfun_table}. \textbf{NOTE:} without ground state partition function
$2 J_0 +1$, included in Table~\ref{nuclei_table}. Total partition function
\eqref{G_kT} is a sum of $2 J_0+1$ and function tabulated below. Results are truncated below $10^{-6}$.}
\begin{tabular}{c|c|c}
Col & Symbol & Description \\
\hline
\hline
1 & No     & position   \\
2 & Symbol & Standard element symbol\\
3 & kT=0.2 & Partition function for $kT=0.2$ MeV \\
4 & kT=0.4 & Partition function for $kT=0.4$ MeV \\
5 & kT=0.6 & Partition function for $kT=0.6$ MeV \\
6 & kT=0.8 & Partition function for $kT=0.8$ MeV \\
7 & kT=1.0 & Partition function for $kT=1.0$ MeV \\
\end{tabular}
\end{table}

\begin{table}
\caption{Symbols used in NSE proton and neutron abundance table, 
main Table~\ref{pn_table}
begins on page \pageref{pn_table}.
}
\begin{tabular}{c|c|c}
Col & Symbol & Description \\
\hline
\hline
1 & kT & Temperature [MeV]\\
2 & $\lg{\rho}$ & base 10 logarithm of the density [g/cm$^3$]\\
3 & $Y_e$ & number of electrons per baryon\\
4 & $X_p$ & abundance of free protons \\
5 & $X_n$ & abundance of free neutrons
\end{tabular}
\end{table}

To calculate all NSE abundance we need basic nuclear data, 
presented in Table~\ref{nuclei_table},
and partition functions, from Table~\ref{partfun_table}.

Using approach presented here, main computational cost is the 
partition function, so use of tabulated version
instead of eq.~\eqref{G_kT} is important.

Detailed description of the algorithm is presented below.
Goal is to calculate abundance $X_k$ of species $k$ for
given temperature $T$, density $\rho$ and $Y_e$:
$$
X_k = NSE(T,\rho, Y_e, k).
$$

\begin{enumerate}
\item from tables of the proton(neutron) abundance we pick up points
surrounding requested $T, \rho, Y_e$; in  case of e.g. trilinear
interpolation these points are 8 corners of a cuboid -- requested point
must be inside or at the edge of the cuboid
\item for all these points we calculate abundance $X_k$ 
from \eqref{nse-abundances}
\item now we have machine-precision accurate abundances
$X_i$ at 8 corners of the cuboid
\item interpolate (trilinear interpolation in the example) 
to get $X_i$ at desired point
\end{enumerate}

We point out again, that we interpolate $X_k$ {\bf NOT} $X_p$ or $X_n$.
$X_k$ must be calculated exactly at grid points. Example implementation
of the algorithm is included in libnse library \cite{libnse}.

\subsection{Implementation notes}

Article deals with interpolation of the functions of three variables.
Despite progress in computer hardware, available memory amount
in particular, it is hard to find sophisticated 3D interpolators.
Therefore trilinear, or mixed bilinear on $T-\rho$ plane and staircase
for $Y_e$, interpolations were used.

A lot of computational time is spend on $X_p^Z X_n^{A-Z}$. Large integer power 
of the floating-point can be computed nearly optimal
using double-exponentiation algorithm, usually included in standard
math libraries. Minor improvements for range of interest can be achieved using 
optimal chain of powers,  C++ template programming or other technique
devoid of \texttt{if} instruction.

We still recommend caution with integer powers of floating point numbers.
For example standard math.h from C does not include integer powers,
GNU Gsl only up to 9, while cmath.h from C++ standard library does.
This cause large variation of the computational time.

Higher-order interpolation
might possibly help to fit procedure into CPU cache memory
thanks to reduced amount of data. However we are also in danger
of overfitting resulting in catastrophic errors, e.g. negative abundances.
If amount of memory is not an issue, linear interpolation is recommended.

Partition function is evaluated using linear interpolation.

\section{Additional numerical data}

Printed tables and results described in the article are meant to be
simple examples of the proposed method.
Extended versions of the tables, custom datasets and numerical library
can be downloaded from \href{http://ribes.if.uj.edu.pl/libnse/}{http://ribes.if.uj.edu.pl/libnse/}
or requested from the author.

\section*{Acknowledgements}

I would like to thank P. Mach, T. Plewa
and K. Kifonidis for valuable discussions and
verification of the NSE results. 
The research was carried out with the supercomputer Deszno purchased thanks
to the financial support of the European Regional Development Fund in the framework
of the Polish Innovation Economy Operational Program (contract no. POIG.
02.01.00-12-023/08).

\bibliographystyle{plain}
\bibliography{NSE_ADNDT}

\clearpage

\section{NSE proton and neutron abundance table}

\begin{center}
\begin{longtable}{ccccccc}
\caption{
\label{nuclei_table}
Nuclei included in NSE
and required nuclear data.}\\
No & Symbol  & Z & N & A & Q & $J_0$ \\
\hline
\endhead
1 & $^{1}$n & 0 & 1 & 1 & 0 & $1/2$\\  
2 & $^{1}$H & 1 & 0 & 1 & 0 & $1/2$\\  
3 & $^{4}$He & 2 & 2 & 4 & 7.0739150 & $0$\\  
4 & $^{12}$C & 6 & 6 & 12 & 7.6801440 & $0$\\  
5 & $^{16}$O & 8 & 8 & 16 & 7.9762060 & $0$\\  
6 & $^{20}$Ne & 10 & 10 & 20 & 8.0322400 & $0$\\  
7 & $^{24}$Mg & 12 & 12 & 24 & 8.2607090 & $0$\\  
8 & $^{28}$Si & 14 & 14 & 28 & 8.4477440 & $0$\\  
9 & $^{32}$S & 16 & 16 & 32 & 8.4931340 & $0$\\  
10 & $^{36}$Ar & 18 & 18 & 36 & 8.5199090 & $0$\\  
11 & $^{40}$Ca & 20 & 20 & 40 & 8.5513010 & $0$\\  
12 & $^{44}$Ti & 22 & 22 & 44 & 8.5335180 & $0$\\  
13 & $^{50}$Ti & 22 & 28 & 50 & 8.7556180 & $0$\\  
14 & $^{48}$Cr & 24 & 24 & 48 & 8.5722100 & $0$\\  
15 & $^{54}$Cr & 24 & 30 & 54 & 8.7779140 & $0$\\  
16 & $^{55}$Cr & 24 & 31 & 55 & 8.7318840 & $3/2$\\  
17 & $^{54}$Mn & 25 & 29 & 54 & 8.7379230 & $3$\\  
18 & $^{55}$Mn & 25 & 30 & 55 & 8.7649880 & $5/2$\\  
19 & $^{56}$Mn & 25 & 31 & 56 & 8.7383000 & $3$\\  
20 & $^{52}$Fe & 26 & 26 & 52 & 8.6095980 & $0$\\  
21 & $^{54}$Fe & 26 & 28 & 54 & 8.7363440 & $0$\\  
22 & $^{55}$Fe & 26 & 29 & 55 & 8.7465600 & $3/2$\\  
23 & $^{56}$Fe & 26 & 30 & 56 & 8.7903230 & $0$\\  
24 & $^{57}$Fe & 26 & 31 & 57 & 8.7702490 & $1/2$\\  
25 & $^{58}$Fe & 26 & 32 & 58 & 8.7922210 & $0$\\  
26 & $^{55}$Co & 27 & 28 & 55 & 8.6695750 & $7/2$\\  
27 & $^{56}$Co & 27 & 29 & 56 & 8.6948170 & $4$\\  
28 & $^{56}$Ni & 28 & 28 & 56 & 8.6427090 & $0$\\  
29 & $^{57}$Ni & 28 & 29 & 57 & 8.6709010 & $3/2$\\  
30 & $^{58}$Ni & 28 & 30 & 58 & 8.7320410 & $0$\\  
31 & $^{60}$Ni & 28 & 32 & 60 & 8.7807570 & $0$\\  
32 & $^{60}$Zn & 30 & 30 & 60 & 8.5832730 & $0$\\  

&&&&
\end{longtable}
\end{center}

\clearpage

\begin{center}
\label{partfun_table}
\begin{longtable}{cccccccc}
\caption{Temperature dependent partition function}\\No & Name &  0.20&  0.40&  0.60&  0.80&  1.00\\ 
\hline
\endhead
3 & $^{4}$He & 0.00 & 0.00 & 0.00 & 0.00 & 0.00\\ 
4 & $^{12}$C & 0.00 & 0.00 & 0.00 &  0.02 &  0.06\\ 
5 & $^{16}$O & 0.00 & 0.00 & 0.00 & 0.00 &  0.03\\ 
6 & $^{20}$Ne & 0.00 &  0.08 &  0.34 &  0.72 &  1.21\\ 
7 & $^{24}$Mg & 0.00 &  0.16 &  0.53 &  1.00 &  1.58\\ 
8 & $^{28}$Si & 0.00 &  0.06 &  0.26 &  0.58 &  1.00\\ 
9 & $^{32}$S & 0.00 &  0.02 &  0.14 &  0.44 &  1.02\\ 
10 & $^{36}$Ar & 0.00 &  0.04 &  0.21 &  0.61 &  1.37\\ 
11 & $^{40}$Ca & 0.00 & 0.00 &  0.04 &  0.24 &  0.87\\ 
12 & $^{44}$Ti &  0.02 &  0.38 &  1.32 &  3.09 &  5.90\\ 
13 & $^{50}$Ti & 0.00 &  0.12 &  0.62 &  1.79 &  4.04\\ 
14 & $^{48}$Cr &  0.12 &  0.85 &  1.98 &  3.60 &  5.95\\ 
15 & $^{54}$Cr &  0.08 &  0.74 &  2.09 &  4.58 &  8.62\\ 
16 & $^{55}$Cr &  1.39 &  5.48 &  11.30 &  18.60 &  27.00\\ 
17 & $^{54}$Mn &  10.90 &  22.30 &  36.70 &  54.70 &  75.20\\ 
18 & $^{55}$Mn &  4.36 &  7.81 &  14.00 &  24.00 &  37.40\\ 
19 & $^{56}$Mn &  15.80 &  33.00 &  50.80 &  70.20 &  90.80\\ 
20 & $^{52}$Fe &  0.07 &  0.63 &  1.49 &  2.70 &  4.46\\ 
21 & $^{54}$Fe & 0.00 &  0.19 &  1.11 &  3.43 &  7.57\\ 
22 & $^{55}$Fe &  0.33 &  2.12 &  6.25 &  13.20 &  22.50\\ 
23 & $^{56}$Fe &  0.07 &  0.68 &  2.01 &  4.78 &  9.60\\ 
24 & $^{57}$Fe &  7.67 &  12.80 &  20.20 &  30.50 &  43.60\\ 
25 & $^{58}$Fe &  0.09 &  0.87 &  2.81 &  6.71 &  13.10\\ 
26 & $^{55}$Co & 0.00 &  0.07 &  0.74 &  2.86 &  7.02\\ 
27 & $^{56}$Co &  4.07 &  10.60 &  18.50 &  27.90 &  38.50\\ 
28 & $^{56}$Ni & 0.00 & 0.00 &  0.08 &  0.32 &  0.86\\ 
29 & $^{57}$Ni &  0.14 &  1.04 &  2.46 &  4.63 &  7.87\\ 
30 & $^{58}$Ni & 0.00 &  0.17 &  0.95 &  2.93 &  6.68\\ 
31 & $^{60}$Ni & 0.00 &  0.25 &  1.31 &  3.89 &  8.48\\ 
32 & $^{60}$Zn &  0.03 &  0.46 &  1.36 &  2.84 &  5.03\\ 
&&&&&&&
\end{longtable}
\end{center}

\clearpage


\begin{center}
\begin{longtable}{|c|c|c||c|c|}
\caption{ 
\label{pn_table}
Proton and neutron abundance tables. } \\
kT  & $\lg{\rho}$ & $Y_e$ & $X_p$ & $X_n$ \\
\hline
\hline
\endfirsthead
\caption[]{ Proton and neutron abundance tables [continued]. } \\
kT  & $\lg{\rho}$ & $Y_e$ & $X_p$ & $X_n$ \\
\hline
\hline
\endhead
0.20&	6&	0.350&	1.4733451757175688e-32&	1.9792286428615166e-01\\
0.20&	8&	0.350&	4.6594595569570535e-37&	1.9791728691182955e-01\\
0.20&	10&	0.350&	1.4734610247732251e-41&	1.9791672869601915e-01\\
0.40&	6&	0.350&	7.6850895802764179e-10&	2.0039781283597258e-01\\
0.40&	8&	0.350&	2.5035658289594687e-14&	1.9822605323795905e-01\\
0.40&	10&	0.350&	7.9454130714268810e-19&	1.9794846194430299e-01\\
0.60&	6&	0.350&	8.6190577876701473e-03&	3.0861905778717863e-01\\
0.60&	8&	0.350&	1.6076815125584695e-06&	2.0253972809389420e-01\\
0.60&	10&	0.350&	5.3242575469851928e-11&	1.9823081609598006e-01\\
0.80&	6&	0.350&	3.4503551381543635e-01&	6.4503551381543633e-01\\
0.80&	8&	0.350&	6.0527987083834160e-03&	3.0605116217065886e-01\\
0.80&	10&	0.350&	5.6832537960862639e-07&	1.9976073216155885e-01\\
1.00&	6&	0.350&	3.4999839054511411e-01&	6.4999839054511410e-01\\
1.00&	8&	0.350&	1.6548297105954407e-01&	4.6548297105954412e-01\\
1.00&	10&	0.350&	1.5943665263060269e-04&	2.1299547811356118e-01\\
0.20&	6&	0.400&	4.5278905162835453e-32&	8.3343373485753311e-02\\
0.20&	8&	0.400&	1.4321164378624826e-36&	8.3334338486210496e-02\\
0.20&	10&	0.400&	4.5288358535831710e-41&	8.3333433860011616e-02\\
0.40&	6&	0.400&	2.2541535853182530e-09&	8.7066391596373041e-02\\
0.40&	8&	0.400&	7.6454369365453428e-14&	8.3829450685805496e-02\\
0.40&	10&	0.400&	2.4404642046330322e-18&	8.3384805016069527e-02\\
0.60&	6&	0.400&	1.3274703748466302e-02&	2.1327470374824983e-01\\
0.60&	8&	0.400&	4.5119825481078112e-06&	9.0877967412558644e-02\\
0.60&	10&	0.400&	1.6254313935731902e-10&	8.3849904241914824e-02\\
0.80&	6&	0.400&	3.9448770859198856e-01&	5.9448770859198852e-01\\
0.80&	8&	0.400&	9.4260257833551960e-03&	2.0942531492521452e-01\\
0.80&	10&	0.400&	1.6763122530124213e-06&	8.6551386142692061e-02\\
1.00&	6&	0.400&	3.9999820882445386e-01&	5.9999820882445387e-01\\
1.00&	8&	0.400&	2.0024502450742615e-01&	4.0024502450742611e-01\\
1.00&	10&	0.400&	3.8713189248855591e-04&	1.0652457407487780e-01\\
0.20&	6&	0.450&	1.2254155814326638e-15&	1.0461956026905567e-14\\
0.20&	8&	0.450&	1.3863583647470302e-17&	1.0941992041239904e-16\\
0.20&	10&	0.450&	1.5862580710493338e-19&	1.1335873063097257e-18\\
0.40&	6&	0.450&	1.3854242348476404e-05&	8.6981272932959097e-05\\
0.40&	8&	0.450&	1.6020742194861760e-07&	8.9771845366122210e-07\\
0.40&	10&	0.450&	1.6200879000788178e-09&	1.0323858209007692e-08\\
0.60&	6&	0.450&	2.3969925356851572e-02&	1.2396992535682635e-01\\
0.60&	8&	0.450&	2.6113737999714160e-04&	3.6480691095702282e-03\\
0.60&	10&	0.450&	5.0716727991727212e-06&	2.4807585164625365e-05\\
0.80&	6&	0.450&	4.4414588253856829e-01&	5.4414588253856822e-01\\
0.80&	8&	0.450&	1.7653064602620443e-02&	1.1765299048431674e-01\\
0.80&	10&	0.450&	2.7450088757524307e-04&	1.5448501064190729e-03\\
1.00&	6&	0.450&	4.4999809512767586e-01&	5.4999809512767583e-01\\
1.00&	8&	0.450&	2.4074257595157450e-01&	3.4074257595157448e-01\\
1.00&	10&	0.450&	2.6369076822919937e-03&	2.3110858589751739e-02\\
0.20&	6&	0.500&	2.1027565926622363e-06&	3.3591508816189998e-23\\
0.20&	8&	0.500&	2.0902933564115093e-07&	3.9833138940328108e-26\\
0.20&	10&	0.500&	2.1502284101490003e-08&	4.5645516281556141e-29\\
0.40&	6&	0.500&	2.4656257836791810e-02&	1.1948909248170071e-07\\
0.40&	8&	0.500&	2.9898277397867462e-03&	1.2462253458017922e-10\\
0.40&	10&	0.500&	2.7224923203959757e-04&	1.6226416771168784e-13\\
0.60&	6&	0.500&	5.5105593212499610e-02&	5.5105593212499276e-02\\
0.60&	8&	0.500&	2.7482483740179343e-02&	6.7168188134800034e-05\\
0.60&	10&	0.500&	6.4480794989162076e-03&	3.9682763842094195e-08\\
0.80&	6&	0.500&	4.9402975852718162e-01&	4.9402975852718162e-01\\
0.80&	8&	0.500&	4.6130285406649901e-02&	4.6130284779868531e-02\\
0.80&	10&	0.500&	2.1479785529572089e-02&	3.4028673055878073e-05\\
1.00&	6&	0.500&	4.9999805645135403e-01&	4.9999805645135403e-01\\
1.00&	8&	0.500&	2.8751054543310173e-01&	2.8751054543310173e-01\\
1.00&	10&	0.500&	3.0258682045056728e-02&	2.9362486076929912e-03\\

&&&&
\end{longtable}
\end{center}

\end{document}